\documentclass{aa}
\usepackage{amssymb}
\usepackage{amsmath}
\usepackage{xspace}
\usepackage{graphicx}

\usepackage{natbib}

\def\gr{$\gamma$-ray}

\def\gr{$\gamma$-ray}
\def\apj{Ap.J.}
\def\apjl{Ap.J.Lett.}
\def\aap{A\&A}
\def\pasp{PASP}
\def\mnras{MNRAS}
\def\sovast{Sov.Astr.Lett.}
\def\araa{ARA\&A}

\begin{document}
\title{Particle acceleration in the vacuum gaps in black hole magnetospheres}

\author{K. Ptitsyna$^{1,2,3}$, A. Neronov$^1$}

\institute{
$^1$ ISDC, Astronomy Department, University of Geneva, Ch. d'Ecogia 16, 1290, Versiox, Switzerland \\
$^2$ Institute for Nuclear Research of the Russian Academy of Sciences,
	60th October Anniversary Prospect 7a, Moscow 117312, Russia \\
$^3$ Physics Department, M.V. Lomonosov Moscow State University, Moscow 119991, Russia}

\abstract
{}
{We consider particle acceleration in vacuum gaps in magnetospheres of black holes powered through  Blandford-Znajek mechanism and embedded into radiatively-inefficient accretion flow (RIAF) environment. In such situation the gap height is limited by the onset of gamma-gamma pair production  on the infrared photons originating from the RIAF.}
{We numerically calculate acceleration and propagation of charged  particles taking into account the detailed structure of electric and magnetic field in the gap and in the entire black hole magnetosphere, radiative energy losses and interactions of \gr s produced by the propagated charged particles with the background radiation field of RIAF.}
{We show that the presence of the vacuum gap  has clear observational signatures. The spectra of emission from gaps embedded into a relatively high luminosity RIAF are dominated by the inverse Compton emission with a sharp, super-exponential cut-off in the very-high-energy gamma-ray band. The cut-off energy  is determined by the properties of the RIAF and is largely independent of the structure of magnetosphere and geometry of the gap. The spectra of the gap residing in  low-luminosity RIAFs are dominated by synchrotron / curvature emission with the spectra extending into  1-100~GeV energy range. We also consider the effect of possible acceleration of protons in the gap and find that  proton energies could reach the ultra-high-energy cosmic ray (UHECR) range only in extremely low luminosity  RIAFs.}
{}


\maketitle

\section{Introduction}

Observations of fast variability of high-energy and very-high-energy \gr\ emission from blazars and radio galaxies \citep{PKS2155_HESS,M87_HESS,M87_HESSMAGIC,IC310_MAGIC,Vovk_Neronov,Tavecchio}
indicate that the region of  particle acceleration and of the associated high-energy \gr\ emission is situated in the vicinity of the supermassive black hole powering these sources. This indirect evidence is supported by the direct measurement of the size of the \gr\ emission region using the effect of gravitational microlensing \citep{neronov15,vovk15}. 

Most of the fast variable sources with compact high-energy \gr\ emission regions belong to the Fanaroff-Riley type I (FR I) radio galaxies  / BL Lac type object classes.  These two source types are different versions of one and the same class of low-luminosity radio galaxies with jets misaligned (for the FR I) or aligned (for BL Lacs) along the line of sight \citep{urry95}. In these sources the central engine of the active galactic nucleus (AGN) is probably to be embedded in a Radiatively-Ineffiecient Accretion Flow (RIAF) which is characterised by moderate matter density and luminosity \citep{narayan98,chiaberge99}.

 Several possible mechanisms of particle acceleration in the compact region near the black hole could be considered. 
 High-energy particle acceleration in the astrophysical environments is conventionally attributed to the shock acceleration mechanism \citep{krymskii77,bell78,drury83}. In the case of radio galaxies, relativistic shocks form in the relativistic jet ejected by the black hole \cite{bridle84}. A shock at the base of the jet close to the black hole could be the source of high-energy particles \cite{marscher08}. Otherwise, the  jet and the RIAF typically generate strong magnetic fields with the energy density reaching the equipartition with the kinetic / gravitational potential energy of the accretion flow. Reconnection of magnetic field lines in the innermost part of the RIAF could lead to the generation of transient electric field accelerating high-energy particles \cite{romanova92,yuan03,yuan14}. 
 Finally, if the black hole is rotating, it works as a unipolar inductor generating large scale electric field in the presence of external magnetic field \cite{wald74}. Such electric field could also accelerate particles which could produce fast variable \gr\ emission \cite{boldt99,levinson00,neronov05,neronov07,krawczynski07,neronov09}. 
 
Rotating black hole could also be the source of the energy for the source activity. Power could be extracted from the black hole rotation e.g. via the Blandford-Znajek process \citep{blandford-znajek}. The Blandford-Znajek process operates in the force-free black hole magnetosphere filled with plasma. However, the plasma is continuously "washed out" from the magnetosphere and needs to be replenished via in-situ generation of free charges. On one hand, this leads to possible generation of the regions with a charge deficit, namely the ``gaps'', where the force-free condition is broken and large scale electric field accelerates particles. On the other hand, it is these gaps, where the charge generation could be achieved. The accelerated particles induce pair production cascades which supply the force-free part of the magnetosphere with charges \citep{blandford-znajek}. This mechanism is identical to that operating in the pulsar magnetospheres \cite{goldreich69,sturrock71,cheng86}. 

The possibility of  existence of charge-starved regions in black hole magnetosphere was discussed by \citet{blandford-znajek,beskin92,levinson00,Levinson_Rieger}.  \citet{beskin92,hirotani98} have shown that the gaps could be considered as the analogs of the outer gaps in pulsar magnetospheres, lying near the surfaces separating areas with opposite space charge densities.	 \citet{krawczynski07,vincent10} have made first attempts to calculate the spectrum of synchrotron and inverse Compton emission from electrons accelerated in the vacuum gap and details of development of the pair cascade induced by the gamma-gamma pair production.
The analysis of the gap properties in particular cases was done by \citet{Levinson_Rieger,tchekhovskoy}. 

In what follows we study observational signatures of operation of vacuum gaps in the magnetospheres of black holes powered by the Blandford-Znajek process. We argue that such observational signatures could be searched for in the signal of radio galaxies and BL Lacs.  In Section \ref{sec:gaps} we discuss the properties of the gaps in black hole magnetospheres using analytical estimates. 
 In section \ref{sec:numeric} we verify the analytical estimates with numerical simulations and compute the spectra of electromagnetic emission from the gaps. Section \ref{sec:discussion} we discuss the results of numerical modelling and  link them with the observations.

\section{Qualitative estimates of parameters of the vacuum gaps}
\label{sec:gaps}

\subsection{Gap location in the magnetosphere}
\label{sec:gap_location}

The electric charge volume density in stationary axisymmetric force-free magnetosphere is given by the Goldreich-Julian density \citep{beskin92}
\begin{equation}
\rho_{GJ}=-\frac{1}{4\pi}\vec{\nabla}\cdot\left(\frac{\Omega_{F}-\omega}{2\pi\alpha}\vec{\nabla}\Psi\right)
\end{equation}
where $\Psi=\Psi(r,\theta)$ is a magnetic flux within the axisymmetric ``magnetic tube'', formed by the rotation of the magnetic field line, passing through the $(r,\ \theta)$ point,  $\Omega_{F}=\Omega_{F}(\Psi)$ is an ``angular velocity''  of the magnetic line, constant on the magnetic flux tubes.  $(r,\theta,\phi)$ is the spherical coordinate system in the Kerr space-time described by the metric 
\begin{eqnarray}
&&ds^2=-\alpha^{2}{dt}^2+g_{jk}(dx^{j}+{\beta}^{j}dt)(dx^{k}+{\beta}^{k} dt)\\
&&g_{rr}=\frac{\rho^{2}}{\Delta}, \ \  g_{\theta\theta}=\rho^{2}, \ \ g_{\phi\phi}={\tilde{\omega}}^{2}, \ \ g_{jk}=0\ \ \mathrm{for}\ \  j \neq k\nonumber\\
&&\alpha=\frac{\rho\sqrt{\Delta}}{\Sigma}, \ \ \beta^r=\beta^{\theta}=0, \ \ \beta_{\phi}=-\omega,\nonumber\\
&&\Delta=r^2+a^2-2Mr, \ \ \rho^2=r^2+a^2{\cos{\theta}}^2, \ \ \omega=\frac{2aMr}{\Sigma},  \nonumber\\
&&\Sigma^2=(r^2+a^2)^2-a^2\Delta\sin{\theta}, \ \ \tilde{\omega}=\frac{\Sigma}{\rho}\sin{\theta},\nonumber
\end{eqnarray}
where $M$ and $a=J/M$, mass and specific angular momentum of a rotating black hole. 
In the following we will always assume that the particle velocities, charge and current densities and the fields components are those measured in the Locally Non-Rotating Frames, LNRF, \citep{bardeen72}.  We use the system of units in which the gravitational constant and the speed of light are unities: $G_N=1; c=1$.

 \begin{figure}
\includegraphics[width=\linewidth]{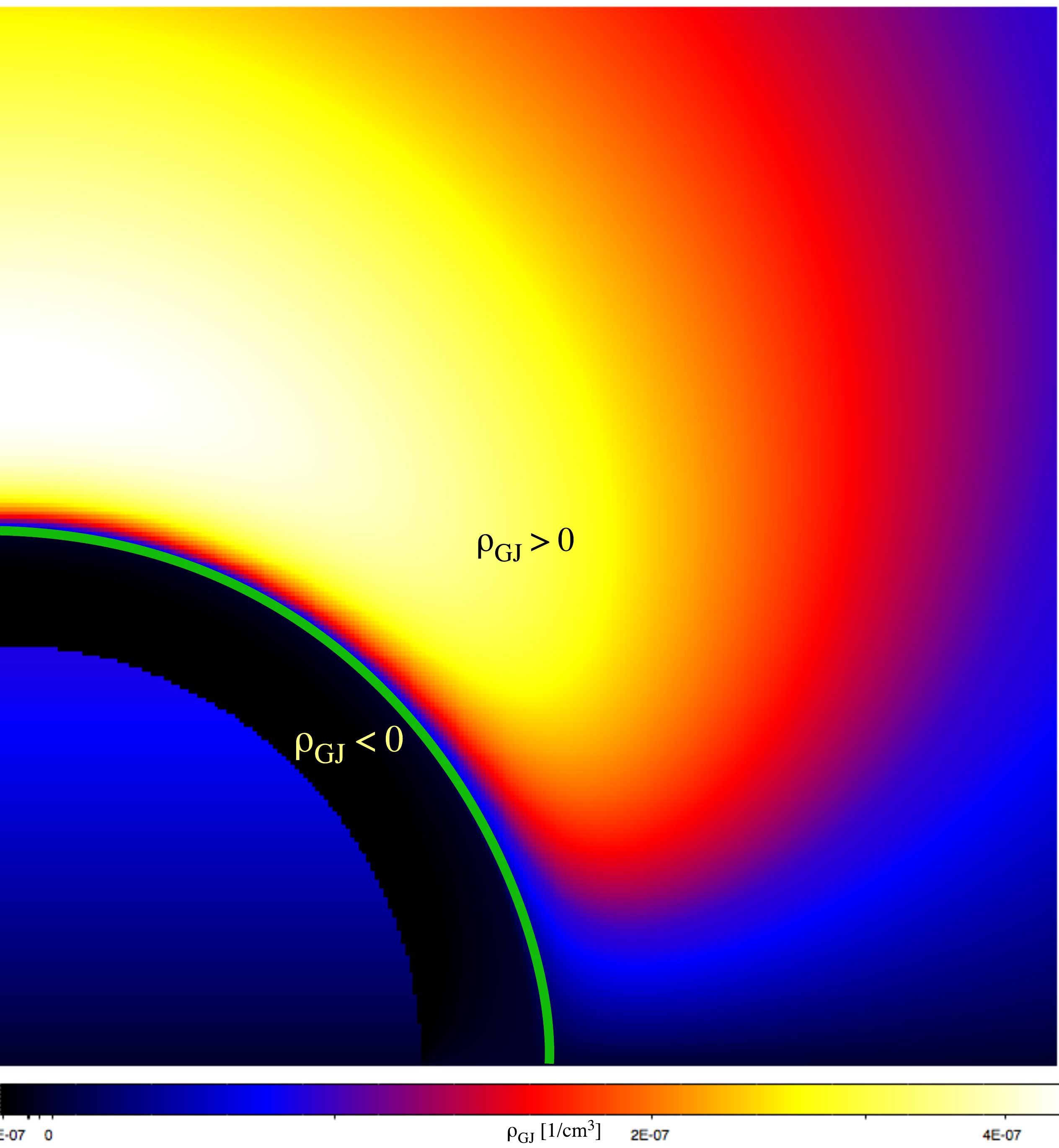}
\caption{\footnotesize{The Gouldreich-Julian charge density in the split-monopole force-free black hole magnetosphere. The black hole rotation moment is $a=0.1M$. The horizon is the boundary of the blue circle in the left bottom corner of the map. The black area is the region with negative charge density. The color scale shows the region of positive values of $\rho(r,\theta)$ with the color scaling logarithmically with $\rho_{GJ}$. Green curve shows the location of the gap at the surface $\rho_{GJ}=0$.  }}
 \label{fig:gj}
 \end{figure}  

Fig \ref{fig:gj} shows the Goldreigh-Julian density map for the black hole force-free magnetosphere with the split-monopole global geometry \citep{blandford-znajek}.   Numerical simulations show that split monopole field,  the simplest analytical model for the global magnetic field configuration in the magnetosphere,  naturally forms in the vicinity of the black hole horizon of a black hole embedded into uniform magnetic field  \citep{komissarov04a,komissarov04,komissarov05}. 
In locations where the charge density is exactly equal to the Goldreigh-Julian density, the electromagnetic field satisfies the force-free condition
\begin{equation}
\vec E+\left[\vec v\times \vec B\right]=0
\end{equation}
where $\vec v$ is the bulk velocity of the plasma filling the magnetosphere and $\vec E,\vec B$ are the electric and magnetic field.

The force-free condition could be violated at locations where peculiarities of the plasma flow do not allow adjustment of the charge balance to provide the Goldreich-Julian charge density. In this case a gap in the force-free magnetosphere could be formed. Such gap forms first of all around the surfaces $\rho_{GJ}=0$ \citep{beskin92}. We will call these surfaces as GJ surfaces. Indeed, inside and outside the GJ surface, there are two surfaces through which charges, both of negative and positive sign, could flow in the only one direction: outwards through the light cylinder and inwards through the horizon. In the absence of continuous injection of particles, the part of the magnetosphere between these two one-way surfaces gets "charge starved". Since this washing out does not depend on the charges sign, the charge deficit occurs at the first place at location of the lowest charge density, i.e. at the GJ surface.

  Following  \citet{beskin92} we use an analytical approximation for the  electric field in the gap
\begin{equation}
\vec{E}=\vec{E}_{\parallel}+\vec{E}_{\perp}
 \end{equation} 
 where $\vec{E}_\perp$ is the poloidal force-free electric field component, perpendicular to the magnetic field and $\vec{E}$ is the component parallel to the magnetic field which satisfies the equation 
 \begin{equation}
 \vec{\nabla}\cdot \vec{E}_{\parallel}=4\pi\left(\rho_e-\rho_{GJ}\right)
 \end{equation}
 with $\rho_e$ being the total charge density. Choosing the $x$ axis along $\vec{B}$ field away from the black hole with $x=0$ on the GJ surface one could find an approximate solution to the above equation
 \begin{equation}\label{f3}
 \vec{E}_{\parallel}=-4\pi\left(\frac{x^2}{2}-\frac{H^2}{8}\right)\frac{d\rho_{GJ}}{dx}\bigg|_{r_{GJ}}.
 \end{equation}   
 which satisfies the boundary condition $E_\parallel=0$ at the boundaries of the gap of the height $x=\pm H/2$, where $H$ is the gap height. 
 
\subsection{Particle acceleration in the gap}

An order-of-magnitude estimate of the energies of particles accelerated in the gap could be found in the following way. 
Adopting an estimate  $\rho_{GJ}\sim \left(B/r\right)\left(a/M\right)$ and $d\rho_{GJ}/dx\bigg|_{r_{GJ}}\sim \left(B/R_{H}^2\right)\left(a/M\right)$, where $R_H=M+\sqrt{M^2-a^2}$ is the  horizon radius, we find the maximal parallel electric field in the gap 
\begin{equation}
E_\parallel\sim \frac{Bh^2a}{M}
\end{equation}
where $h=H/R_H$. The maximal acceleration rate $d{\cal E}_+/dt$ is 
 \begin{equation}
 \label{f3i}
 \frac{d{\cal{E}}_{+}}{dt}\sim \frac{e Bh^2a}{M},
 \end{equation}
where $e$ is the particle charge.

If the rate of energy losses due to the synchrotron/curvature and inverse Compton emission is much lower than the acceleration rate (\ref{f3i}), particles are accelerated in the gap to the maximal energy 
\begin{equation}
{\cal{E}}\sim \frac{eBR_Hh^3a}{M}\sim 10^{15}\left[\frac{h}{0.01}\right]^3\left[\frac{B}{10^4\mbox{ G}}\right]\left[\frac{M}{10^9M_\odot}\right]\biggl[\frac{a}{M}\biggr]\mbox{ eV}
\end{equation}
This is nearly always the case for protons. To the contrary, electrons suffer from much stronger energy losses. Their maximal energies are determined by the balance of the energy gain rate (\ref{f3i}) and the energy loss. The loss rate is dominated by the synchrotron / curvature radiation loss for the low-luminosity RIAF:
\begin{equation}
\frac{d{\cal E}}{dt}\sim -\frac{e^2{\cal E}^4}{m_e^4R^2}\sim -\frac{e^4{\cal E}^2B_\bot^2}{m_e^4}
\end{equation}
with $R$ ranging between $R\sim R_H$ for the curvature radiation and $R={\cal E}/eB_\bot$ for the synchrotron radiation in magnetic field with the component  $B_\bot$ normal to the particle velocity. 
The maximal energies attainable by electrons are then
\begin{eqnarray}
&&{\cal E}\sim \frac{h^{1/2}B^{1/4}R_H^{1/2}a^{1/4}}{e^{1/4}M^{1/4}}m_e\sim \\ &&2\times 10^{15}\left[\frac{h}{0.01}\right]^{1/2}\left[\frac{B}{10^4\mbox{ G}}\right]^{1/4}\left[\frac{M}{10^9M_\odot}\right]^{1/2}\biggl[\frac{a}{M}\biggr]^{1/4}\mbox{eV}\nonumber
\end{eqnarray}
for the curvature radiation and 
\begin{eqnarray}
&&{\cal E}\sim \frac{hm_e^2B^{1/2}a^{1/2}}{e^{3/2}B_{\bot}M^{1/2}}\sim \\ &&4\times 10^{13}\left[\frac{h}{0.01}\right]\left[\frac{B}{10^4\mbox{ G}}\right]^{1/2}\left[\frac{B_\bot}{1\mbox{ G}}\right]^{-1}\biggl[\frac{a}{M}\biggr]^{1/2}\mbox{eV}\nonumber
\end{eqnarray}
for the dominant synchrotron radiation loss (where we have assumed $B_\bot\sim E_\parallel$ for the numerical estimate).

High-luminosity RIAF leads to the dominance of the inverse Compton energy loss which scales as 
\begin{equation}
\frac{d{\cal E}}{dt}\sim -\frac{\sigma_TU_{rad}{\cal E}^2}{m_e^2}
\end{equation}
in the Thomson regime. This loss limits electron energies to 
\begin{eqnarray}
&&{\cal E}\sim \frac{(4\pi)^{1/2}e^{1/2}B^{1/2}hm_eR_{ir}a^{1/2}}{\sigma_T^{1/2}L^{1/2}M^{1/2}}\sim\nonumber \\ &&7\times 10^{13}\left[\frac{h}{0.01}\right]\left[\frac{B}{10^4\mbox{ G}}\right]^{1/2}\left[\frac{M}{10^9M_\odot}\right] \\ &&\ \ \ \ \ \ \ \ \ \ \ \ \ \ \ \ \ \ \ \ \ \ \ \ \ \ \ \ \ \ \ \ \ \left[\frac{L}{10^{42}\mbox{ erg/s}}\right]^{-1/2}\biggl[\frac{a}{M}\biggr]^{1/2}\mbox{eV}\nonumber
\end{eqnarray}
where $R_{ir}\sim 10R_H$ is the characteristic size of the infrared source in RIAF.


{The number of electrons in the gap $N_{e}$ is limited by the Goldreich-Julian density $\rho_{GJ}$. The gap volume is estimated as $HR_H^2$ and we have $N_{e}=(\rho_{GJ}/e)hR_H^3$. 
Thus the maximum total luminosity of the gap $L_{gap}$ due to the electrons synchrotron/curvature or inverse Compton radiation in the energy gain - energy loss balanced regime is given by:}
\begin{eqnarray}\label{fL}
 &&L_{gap}\sim N_{e}(d{\cal{E}}/dt)\sim \\ &&3\times10^{40}\biggl[\frac{a}{0.1M}\biggr]^2\left[\frac{h}{0.1}\right]^3\left[\frac{B}{10^3\mbox{G}}\right]^2\left[\frac{M}{10^9M_{\odot}}\right]^2 \mbox{erg/s}\nonumber 
\end{eqnarray}

\subsection{Gamma-gamma pair production in the gap}

The height of the gap is limited by the onset of the electron-positron pair production in interactions of the inverse Compton / curvature / synchrotron photons with the infrared radiation from the RIAF. The threshold \gr\ energy for the pair production on the infrared photons of the energy $\epsilon_{ir}$ is 
\begin{equation}
E_{\gamma,thr}=\frac{m_e^2}{\epsilon_{ir}}\simeq 3\left[\frac{\epsilon_{ir}}{0.1\mbox{ eV}}\right]^{-1}\mbox{ TeV}
\end{equation}
The energy of the curvature  
\begin{equation}
E_\gamma=\frac{{\cal E}^3}{m_e^3R}\sim 10^9\left[\frac{\cal E}{10^{15}\mbox{ eV}}\right]^3
\left[\frac{M}{10^9M_\odot}\right]^{-1}\mbox{ eV}
\end{equation}
and synchrotron 
\begin{equation}
E_\gamma=\frac{eB_\bot{\cal E}^2}{m_e^3}\sim 10^7\left[\frac{B_\bot}{1\mbox{ G}}\right]\left[\frac{\cal E}{10^{13}\mbox{ eV}}\right]^2\mbox{ eV}
\end{equation}
photons typically does not reach the pair production threshold. To the contrary, the energy of the inverse Compton photons

$$ E_\gamma\simeq \begin{cases}
{\cal E}^2/(\epsilon_{ir} m_e^2), &\ \ \  {\cal E}\ll E_{\gamma,thr}\\
{\cal E}, &\ \ \  {\cal E}\gtrsim E_{\gamma,thr}
\end{cases}$$

The pair production process can limit the height of the gap only if the potential difference in the gap is sufficient for the acceleration of electrons to the energy at which the inverse Compton \gr s start to produce pairs:
  \begin{eqnarray}
  \label{f4}
 && h\gtrsim h_{*,acc}\sim \left(\frac{m_e^2}{e\epsilon_{ir}BR_H}\right)^{1/3}\sim \\ &&10^{-3}\left[\frac{\epsilon_{ir}}{0.1\mbox{ eV}}\right]^{-1/3}\left[\frac{B}{10^{4}\mbox{ G}}\right]^{-1/3}\left[\frac{M}{10^{9}M_\odot}\right]^{-1/3}\nonumber
  \end{eqnarray}
If the energies of electrons are limited by the balance between the acceleration rate and inverse Compton loss rate, the minimum gap height necessary for the onset of the pair production is
\begin{eqnarray}
 && h\gtrsim h_{*,ic}\sim\left(\frac{\sigma_TLm_e^2}{4\pi e\epsilon_{ir}^2BR_{ir}^2}\right)^{1/2}\simeq 4\times 10^{-4}\\ &&
\left[\frac{L}{10^{42}\mbox{ erg/s}}\right]^{1/2}\left[\frac{B}{10^4\mbox{ G}}\right]^{-1/2}\left[\frac{M}{10^9M_\odot}\right]^{-1}\left[\frac{\epsilon_{ir}}{0.1\mbox{ eV}}\right]^{-1}\nonumber
 \end{eqnarray}

The pair production occurs on distance scale of the order of the \gr\ mean free path of \gr s
\begin{eqnarray}
&&\lambda_{\gamma\gamma}=\frac{1}{\sigma_{\gamma\gamma}n_{ph}}=\frac{4\pi R_{ir}^2\epsilon_{ir}}{\sigma_{\gamma\gamma}L}\simeq \\ &&2\times 10^{-2}R_H\left[\frac{L}{10^{42}\mbox{ erg/s}}\right]^{-1}\left[\frac{M}{10^9M_\odot}\right]\left[\frac{\epsilon_{ir}}{0.1\mbox{ eV}}\right]\nonumber
\end{eqnarray}
which could become comparable to the distance on which electrons acquire sufficient energy for the pair production, $h_{*,acc}R_H, h_{*,ic}R_H$, if the luminosity of RIAF is low. In general, the gap height could be estimated as
\begin{equation}
h\simeq \min\left(h_{*,acc}, h_{*,ic}\right)+\lambda_{\gamma\gamma}/R_H
\end{equation}

\begin{figure}[h!]
\includegraphics[width=\linewidth]{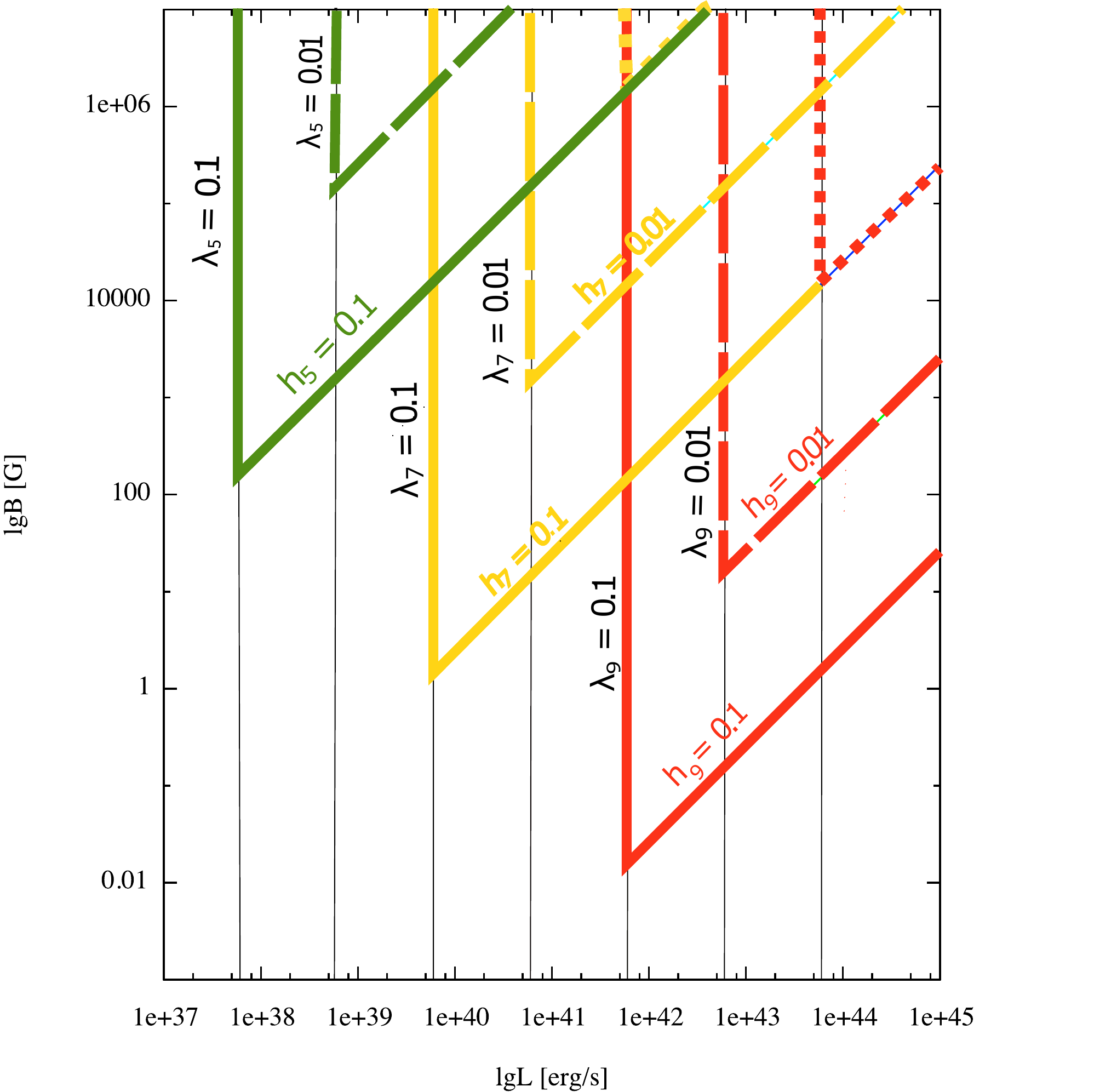}
\caption{Height of the gap as a function of $L,B$ for a range of black hole masses (from left to right): $M=3\times 10^5M_\odot$  (green) $M=3\times 10^7M_\odot$ (yellow) and $M=3\times 10^9M_\odot$ (red).Gap height $h$ and mean free path $\lambda$ are in the units of $R_H$.}
\label{fig:LBdiagram}
\end{figure}

The gap height is determined mostly by $h_{*,ic}$ in the high-luminosity RIAF regime. In this situation the constant gap height lines in the parameter space $L,B$ are the lines $L\propto B$, shown in Fig. \ref{fig:LBdiagram}.  If the inverse Compton energy loss is negligible, the gap height $h_{*,acc}$ is determined by the minimal necessary potential difference in the gap needed to for the onset of the gamma-gamma pair production. $h_{*,acc}$ does not depend on $L$. This regime is reached at the tip of the "wedge" regions  in Fig.  \ref{fig:LBdiagram}. In the regime of small $L$ the mean free path of photons becomes large. This influences the gap height in magnetic-field-independent way. For each pre-defined gap height, there is a minimal "threshold" luminosity scale at which the photon mean free path becomes much larger than the height. This threshold value is shown as a set of vertical lines in Fig. \ref{fig:LBdiagram}. 

Overall, qualitative considerations presented above show that the constant gap height curves form a set of  "wedges"  in the $L,B$ parameter space. For a given magnetic field strength the gap "opens" (becomes large) either when the luminosity of RIAF is too low or it is too high. The infrared photon density in the low luminosity RIAF is not sufficient for the interception of the high-energy \gr s which could produce electron-positron pairs. Strong inverse Compton energy loss in the  high-luminosity RIAF precludes particle acceleration to the energies at which they are capable to produce \gr s which could produce pairs in interactions with the infrared photons. 

\subsection{Gap stability}

Pair production process leads to the deposition of the plasma around the gap location. The initial assumption which was used to calculate the profile of the parallel electric field strength in the gap (\ref{f3}) is that the charge density reaches the Goldreigh-Julian value exactly at the boundary of the gap. 

This assumption could be verified with a more detailed understanding of the process of particle acceleration followed by the pair production. The gamma-gamma pair production homogeneously deposits electron-positron pairs on the distance scale $\lambda_{\gamma\gamma}$. This indicates that the characteristic distance scale on which the charge density and the parallel component of the electric field in the gap vary is $\lambda_{\gamma\gamma}$. It is this distance scale which should be identified with $H$ considered in section \ref{sec:gap_location}.  This identification could clearly be done in the situation when $h_{*,acc},h_{*,ic}\ll \lambda_{\gamma\gamma}/R_H$.  

In the regime of high luminosity, the mean free path of \gr s which are able to produce pairs in interactions with the bulk of the infrared photons of RIAF is much shorter than the distance on which electrons are able to gain the energy sufficient for the production of such \gr s. If such \gr s would be the only \gr s capable of pair production, the configuration would be unstable. Indeed, the gap which initially opens up to the height $h_{*,ic}$ would shrink immediately after opening down the $h\sim \lambda_{\gamma\gamma}\ll h_{*,ic}$ thus shutting down the pair production process. This instability is, however, naturally avoided because of the presence of the broad energy distribution of the RIAF's photons. 

Still, also in this situation the gap height, which is about the electron acceleration length, is also about the \gr\ mean free path. If the spectrum of synchrotron infrared emission from the RIAF emits mostly in the energy band $\epsilon_{ir}$, there are still infrared photons of higher energies, even if the spectrum has an exponential higher energy cut-off. Presence of these higher energy photons provides a possibility for the \gr s with energies somewhat lower than $E_{\gamma,thr}$ to interact via the pair production. Much lower density of the higher energy infrared photons leads to larger mean free path of the lower energy \gr s. There exists an energy at which the mean free path of the \gr s is comparable to the height scale $h_{*,ic}$. Electron acceleration to this energy (i.e. somewhat lower than ${\cal E}=E_{\gamma,thr}$) provides the stable gap configuration in which the charge density in the gap varies on the distance scale $h\sim h_{*,ic}\sim \lambda_{\gamma\gamma}/R_H$. 

A realistic RIAF spectrum consists of three components: synchrotron, inverse Compton and Bremsstrahlung  emission from electrons in the accretion flow \cite{narayan98,yuan14}. Close to the black hole horizon, where heating of electrons by protons is most efficient, the energies of electrons reach $10-100$~MeV and the synchrotron emission from such electrons is in the infrared range. The inverse Compton scattering of the synchrotron photons boosts the photon energies into the X-ray range. The high-energy \gr s could produce pairs not only in interactions with the exponentially suppressed high-energy tail of the synchrotron spectrum of the RIAF, but also with the inverse Compton photons. We leave the detailed modelling of the imprint of the RIAF spectrum and spatial properties for future work.

\subsection{Proton acceleration in the gap}

Contrary to electrons, protons accelerated in the gap do not suffer from the severe energy losses which limit their energy. They could be accelerated to much higher energies than electrons. An estimate of the energies of protons is given by the total potential difference between the gap boundaries, ${\cal E}_p\sim eBR_H h^3$. 

In the low luminosity RIAF regime the gap height is determined by the mean free path of \gr s which produce pairs in interactions with the bulk of the infrared emission from RIAF.  $h\sim \lambda_{\gamma\gamma}/R_H$. In this case the maximal proton energies reach 
\begin{eqnarray}
&&{\cal E}_p\sim \frac{(4\pi)^3eBR_{ir}^6\epsilon_{ir}^3}{\sigma_{\gamma\gamma}^3L^3R_H^2}\simeq 6\times 10^{21}\\ &&
\left[\frac{L}{10^{40}\mbox{ erg/s}}\right]^{-3}\left[\frac{B}{10^{4}\mbox{ G}}\right]\left[\frac{\epsilon_{ir}}{0.1\mbox{ eV}}\right]^3\left[\frac{M}{10^9M_\odot}\right]^4\mbox{eV}\nonumber
\end{eqnarray}
One could see that in this regime proton energies typically do not reach the UHECR range unless the luminosity of RIAF is very low, $L\lesssim 10^{40}$~erg/s. 

In the high luminosity regime, the gap height is estimated as $h\sim h_{*,ic}$ and the maximal energies of protons are
\begin{eqnarray}
&&{\cal E}_p\sim eBR_H\left(\frac{\sigma_TLm_e^2}{4\pi e \epsilon_{ir}^2BR_{ir}^2}\right)^{3/2}
\simeq  5\times 10^{13}\\ &&\left[\frac{L}{10^{44}\mbox{ erg/s}}\right]^{3/2}\left[\frac{B}{10^4\mbox{ G}}\right]^{-1/2}\left[\frac{M}{10^9M_\odot}\right]^{-2}\left[\frac{\epsilon_{ir}}{0.1\mbox{ eV}}\right]^{-3}\mbox{ eV}\nonumber
\end{eqnarray}
 \begin{figure}[h!]
 	\includegraphics[width=\linewidth]{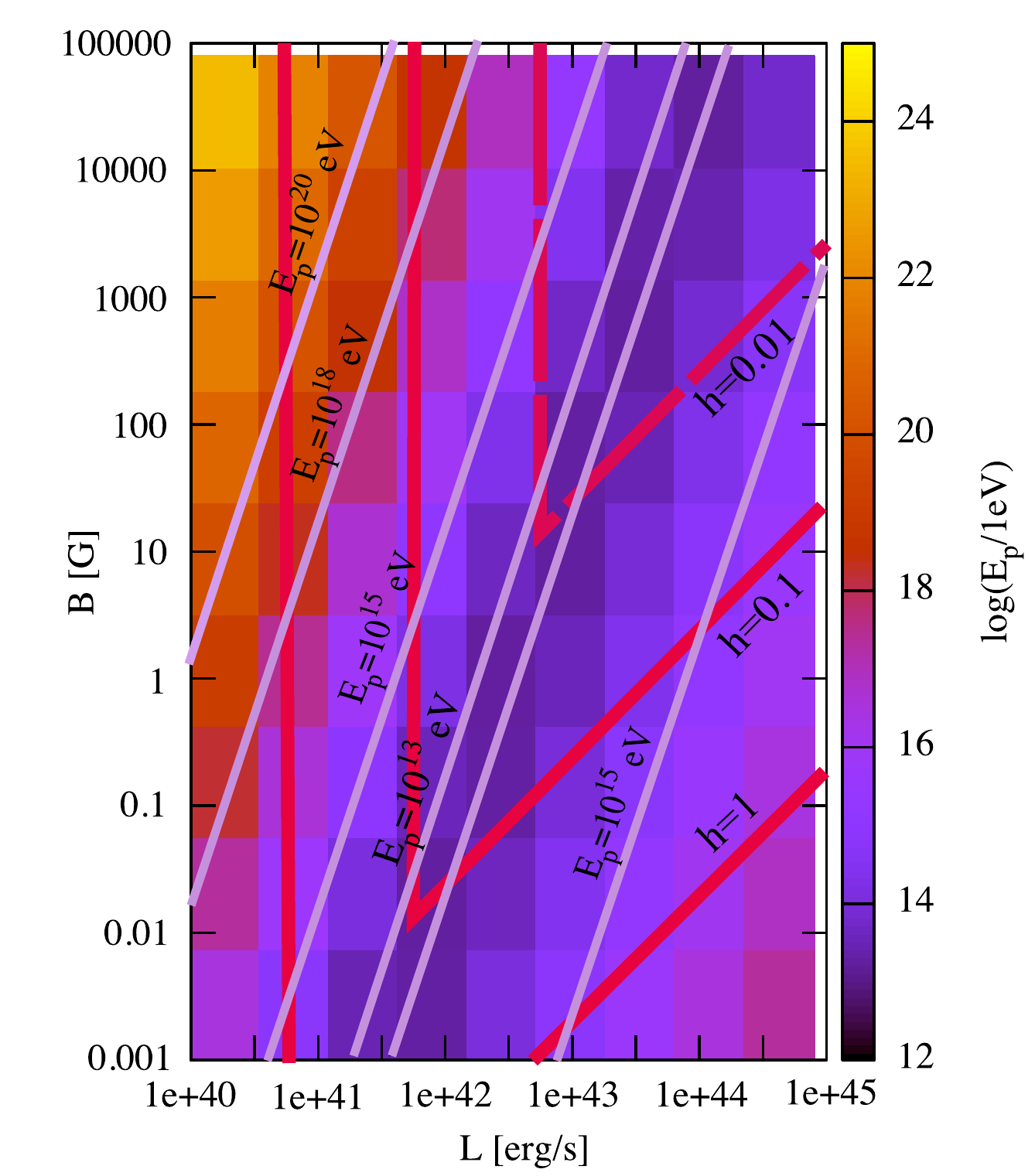}
 	\caption{The protons energy, passing the whole potential difference in the gap, shown as a colormap and its levels in the $(L, B)$ parameter space in logarithmic scale for the black hole mass $M=3\times 10^{9} M_{\odot}$. The dark diagonal valley corresponds to the minimal energy $\sim 10^{13}\mathrm{eV}$ of the protons able to pass the whole gaps potential. This minimal energy does not depend on the black hole mass. The energies of about $\geq 10^{19}\mathrm{eV}$ are reachable only for the ``big'' opened gaps with a height of about $H> 0.1R_{H}$.}
 	\label{protons-analytical}
 \end{figure}
Such gaps could not work as UHECR accelerators. The lines of the constant protons energy, passing the whole potential difference in the gap, in the $(L, B)$ parameter space are shown on the Fig. \ref{protons-analytical}.
%
 
  \section{Numerical modelling}
  \label{sec:numeric}
  
To verify the qualitative arguments presented in the previous sections, we perform Monte-Carlo simulations of propagation of electrons and protons in the split monopole magnetospheres of black holes powered through the Blandford-Znajek process \citep{blandford-znajek}. We consider only slowly rotating black holes with $a=0.1M$. 

Charged particles (electrons or protons) are injected at different locations in the vacuum gap. Particle trajectories inside and outside the gap are calculated solving the equations of motion in the background magnetic and gravitational fields, following the methodology of Ref. \citep{neronov09}. Particles propagating in the force-free part of the magnetosphere and in the gap emit curvature and inverse Compton radiation. The power of this radiation is taken into account to calculate the synchrotron / curvature and inverse Compton energy losses which contribute to the radiative friction force. The inverse Compton energy loss in the Thomson and Klein-Nishina regimes  is calculated based on a realistic emission spectrum of the synchrotron component of RIAF 
\begin{equation}\label{f16}
\frac{dn_{ir}}{d\epsilon}\propto \epsilon^{-1}\exp\left(-\frac{\epsilon}{\epsilon_{ir}}\right)
\end{equation}
with a high energy cut-off at $\epsilon=\epsilon_{ir}$. The overall normalisation of the synchrotron spectrum is determined by the assumed luminosity of RIAF, $L$. The size of the synchrotron emission region is assumed to be $R_{ir}=10R_H$.   Calculation of the spectra of radiation from the propagated charged particles take into account the Doppler and gravitational redshift effects. 

\begin{figure}
\includegraphics[width=\linewidth]{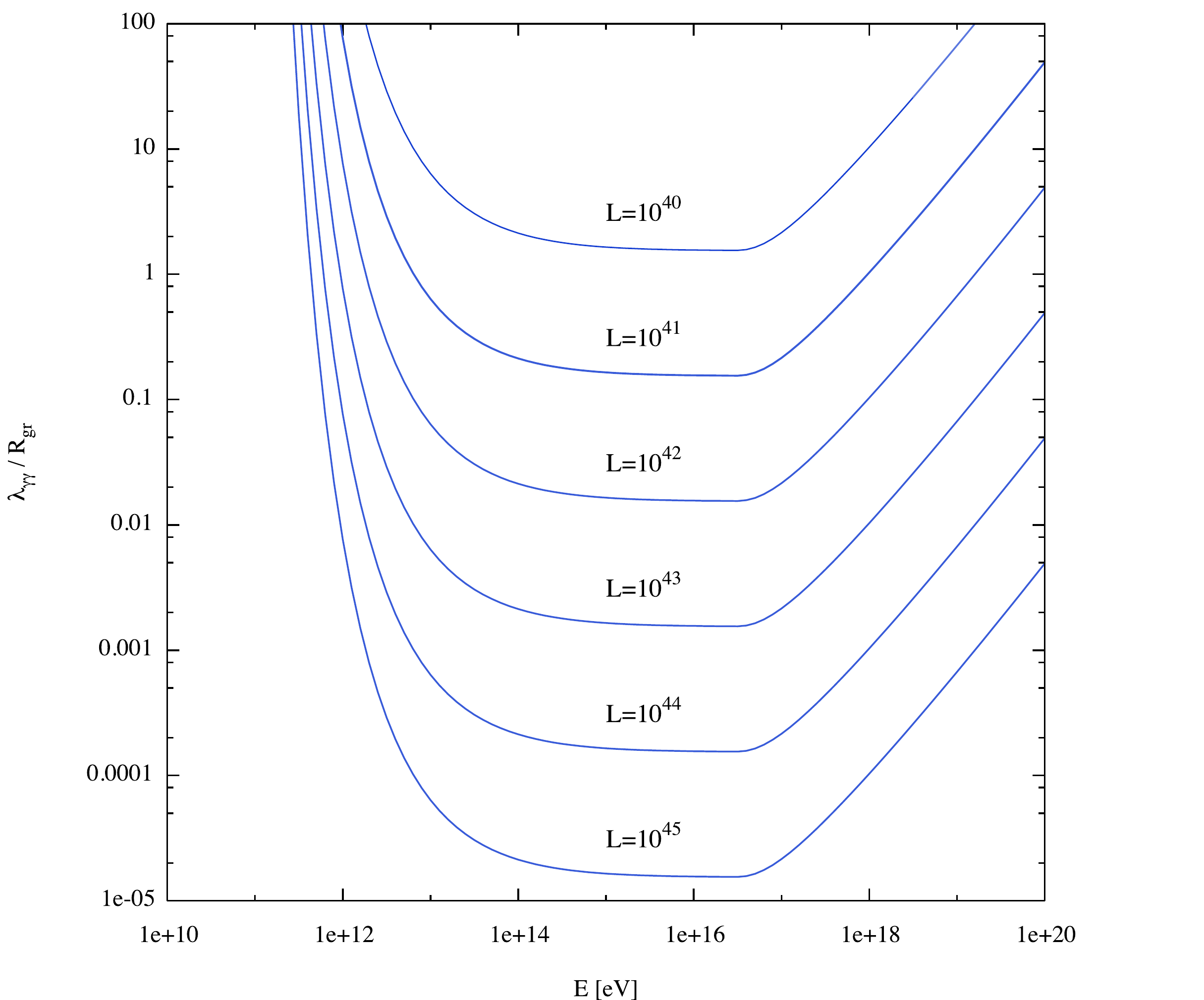}
 \caption{Mean free path of \gr s of energy $E_{\gamma}$ propagating through the photon background of the RIAFs of different luminosities with a model spectrum (\ref{f16}). Black hole mass $M=3\times 10^9 M_{\odot}$.}
 \label{fig:mfp}
 \end{figure}

The gap height is calculated in a self-consistent way as a boundary of the pair production region. For each photon emitted by the propagated charged particle,  we calculate its mean free path  in the background radiation field
\begin{equation}
\lambda_{\gamma\gamma}^{-1}(E_{\gamma})=\int\sigma_{\gamma\gamma}(E_{\gamma},{\epsilon})\frac{dn_{ir}(\epsilon)}{d\epsilon}d\epsilon
\end{equation}
(see Fig. \ref{fig:mfp})
and check if the photon is absorbed within the calculation region. For the photons which are absorbed via pair production, the distances of the pair production points are recorded. The gap height is determined by the locations of the pair production points by the highest energy photons with the shortest mean free path. 
The height of the gap determined in this way is a function of the latitude $\theta$.  
 
\begin{figure}
\includegraphics[width=\linewidth]{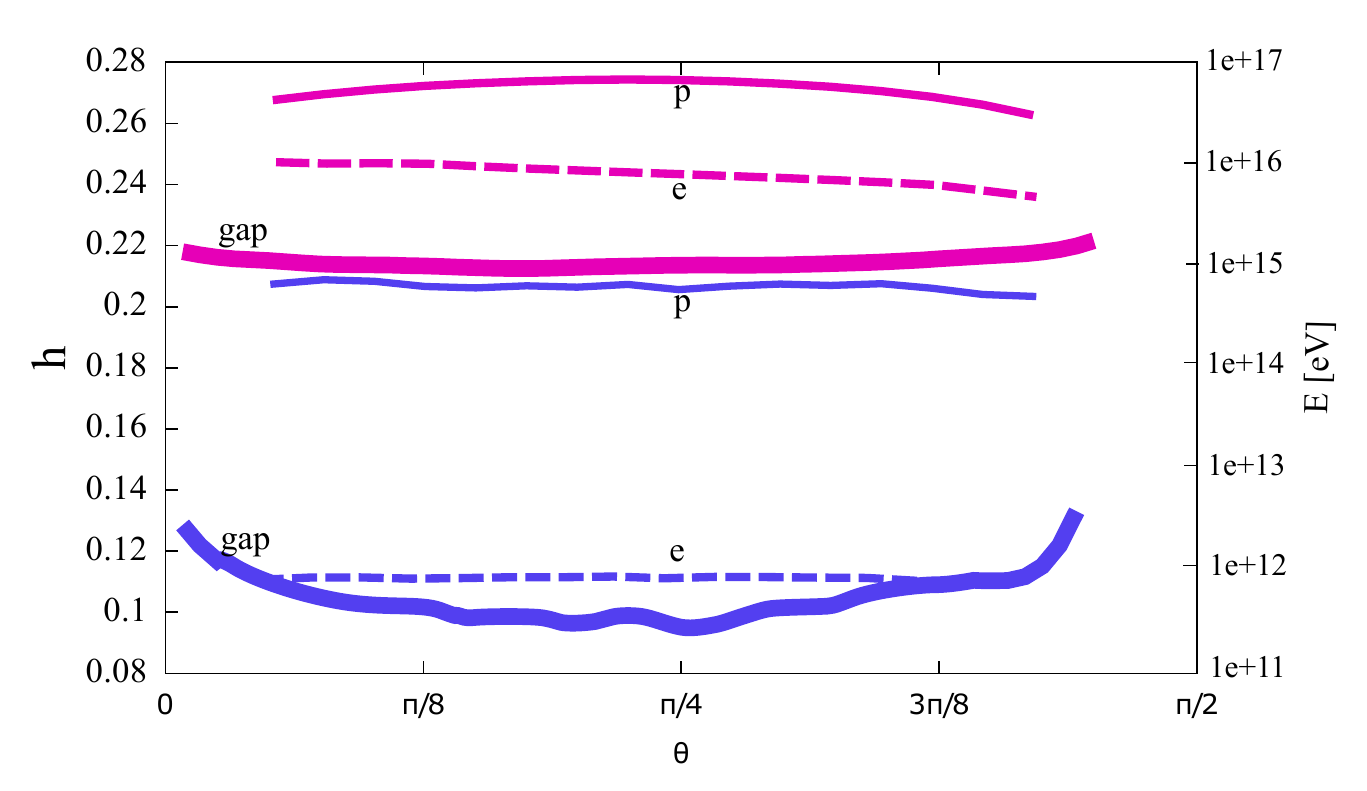}
 \caption{Height of the gap (thick solid curves), proton (thin solid curves) and mean electron (dashed curves) energies in the gap as a function of the latitude $\theta$ for a black hole of the mass $M=3\times10^9M_\odot$, embedded in RIAF with ``low'' luminosity $L=8\times10^{40} \mathrm{erg/s}$, magnetic field $B=10^3\mathrm{G}$ (red curves) and ``high'' luminosity $L=5\times 10^{43}\mathrm{erg/s}$, $B=10^2\mathrm{G}$ (blue curves). Electron and proton energies are almost independent on the latitude $\theta$.}
 \label{fig:gap_height}
 \end{figure}
 
Fig. \ref{fig:gap_height} shows  the dependence of the height of the gap on the  lattitude coordinate $\theta$.
 One could see that the height is almost $\theta$-independent.  One could use the height measurement at a reference
 latitude, e.g. $\theta=45^\circ$ to characterise the height dependence on other parameters of the system. 
 
\begin{figure}
\includegraphics[width=\linewidth]{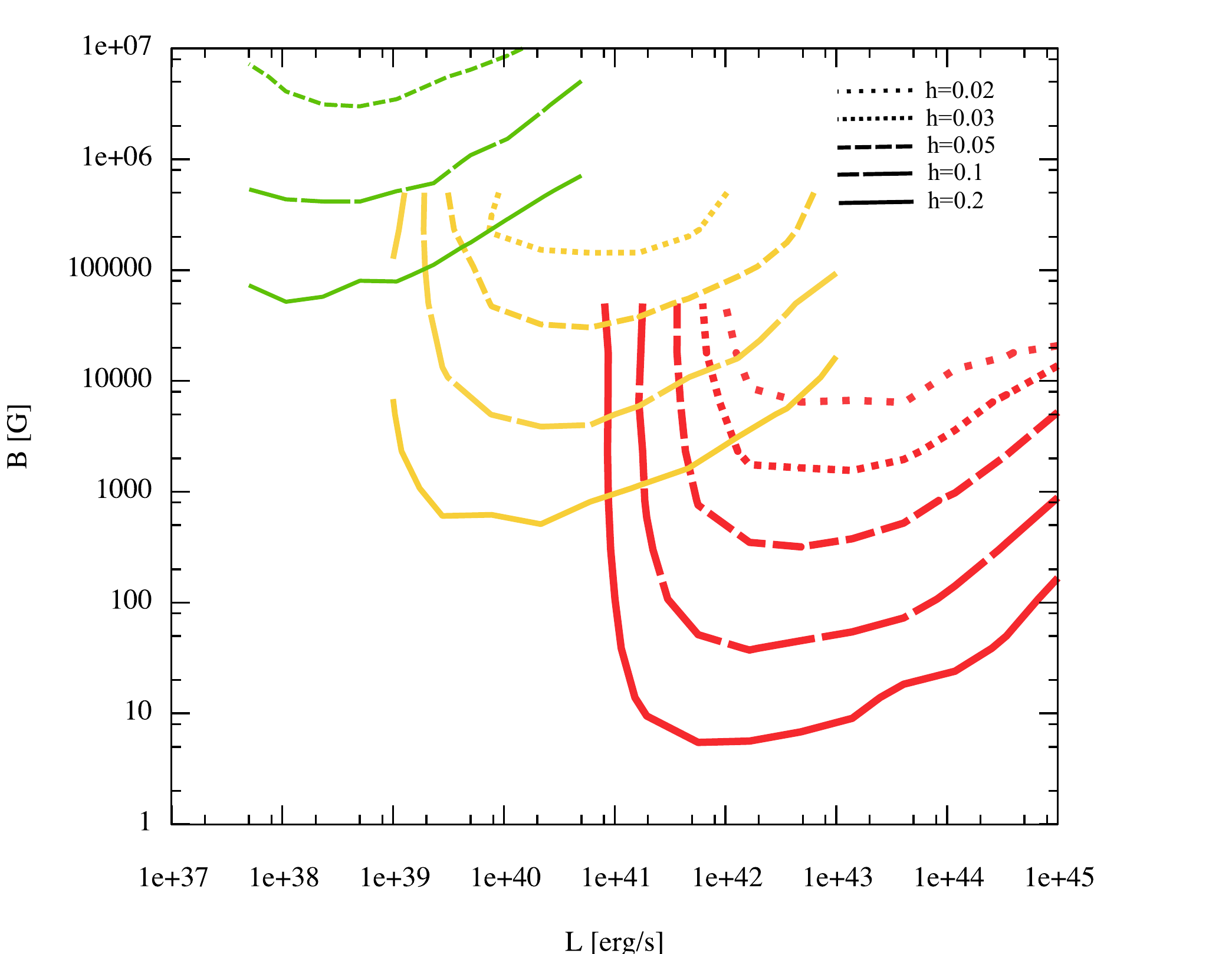}
 \caption{Gap height as a function of RIAF luminosity $L$ and magnetic field $B$ for a range of black hole masses (from left to right): $M=3\times 10^5M_{\odot}$ (green), $M=3\times 10^7M_{\odot}$ (yellow) and $M=3\times 10^9M_{\odot}$ (red).}
 \label{fig:LBdiagram_sim}
 \end{figure}

 Fig. \ref{fig:LBdiagram_sim} shows the gap height at $\theta=45^\circ$ as a function of the RIAF luminosity $L$ and magnetic field $B$  for several black hole masses. One could see that the numerically calculated curves generally agree with the qualitative curves of Fig. \ref{fig:LBdiagram}. The main difference is the absence of the sharp "edge" like feature at $h_{*,acc}=\lambda_{\gamma\gamma}$, which is replaced by a smooth turnover which interpolates between the low-luminosity and high-luminosity asymptotes. The  gap height gets smaller with increasing magnetic field and luminosity.  The gap completely opens (its size becomes large) below certain threshold luminosity, which depends on the black hole mass. The calculaitons shown in Fig. \ref{fig:LBdiagram_sim} extend up to 0.01 Eddington luminosity $L_{Edd}=10^{47}\left[M/10^9M_\odot\right]\mbox{ erg/s}$. This might be somewhat higher than the range of luminosities at which the RIAF model of accretion flow is valid.
 
We find that the observational appearance of the gap  depends on the level of the luminosity of the accretion flow. Taking this into account, we divide the luminosity range onto "low", "intermediate" and "high", depending on the part of the Fig. \ref{fig:LBdiagram_sim} curve at which the system parameters fit. The low-luminosity regime corresponds to the sharply rising (near vertical) part of the constant gap height curves, the "intermediate" regime is in the horizontal plateau range and the "high-luminosity" regime corresponds to the rising part of the curves at high $L$ values.     
 
\begin{figure}
\includegraphics[width=\linewidth]{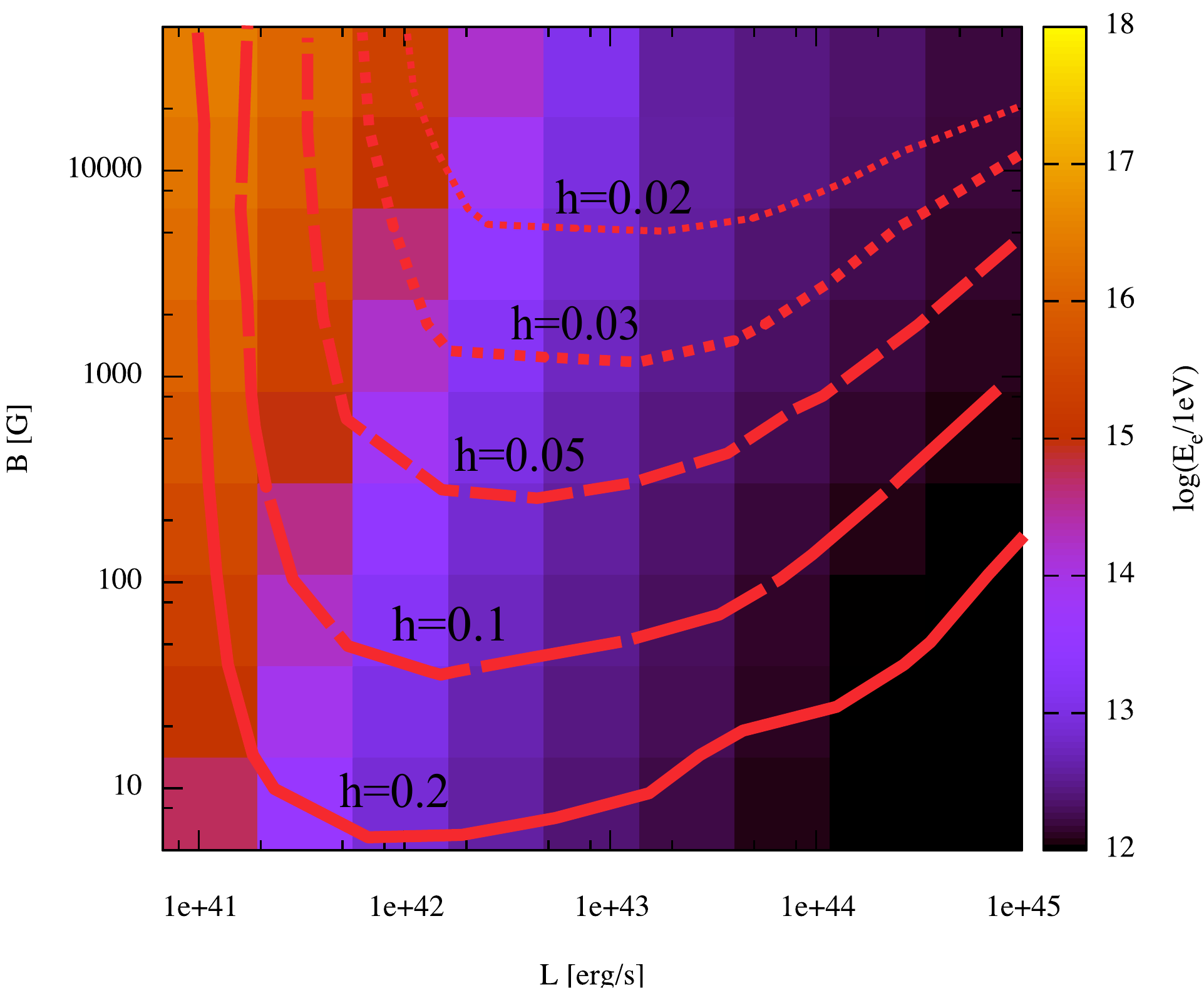}
 \caption{Maximal energies of electrons accelerated in the gap as a function of $L,B$. The black hole mass is $M=3\times 10^9M_\odot$. }
 \label{fig:electron_energy}
 \end{figure}

The maximal energies of electrons in the gap are shown as a function of $L,B$ in Fig. \ref{fig:electron_energy}. The energies increase with the decrease of the luminosity and increase of the magnetic field, as expected from the qualitative arguments presented above. 
  
\begin{figure}
 \includegraphics[width=\linewidth]{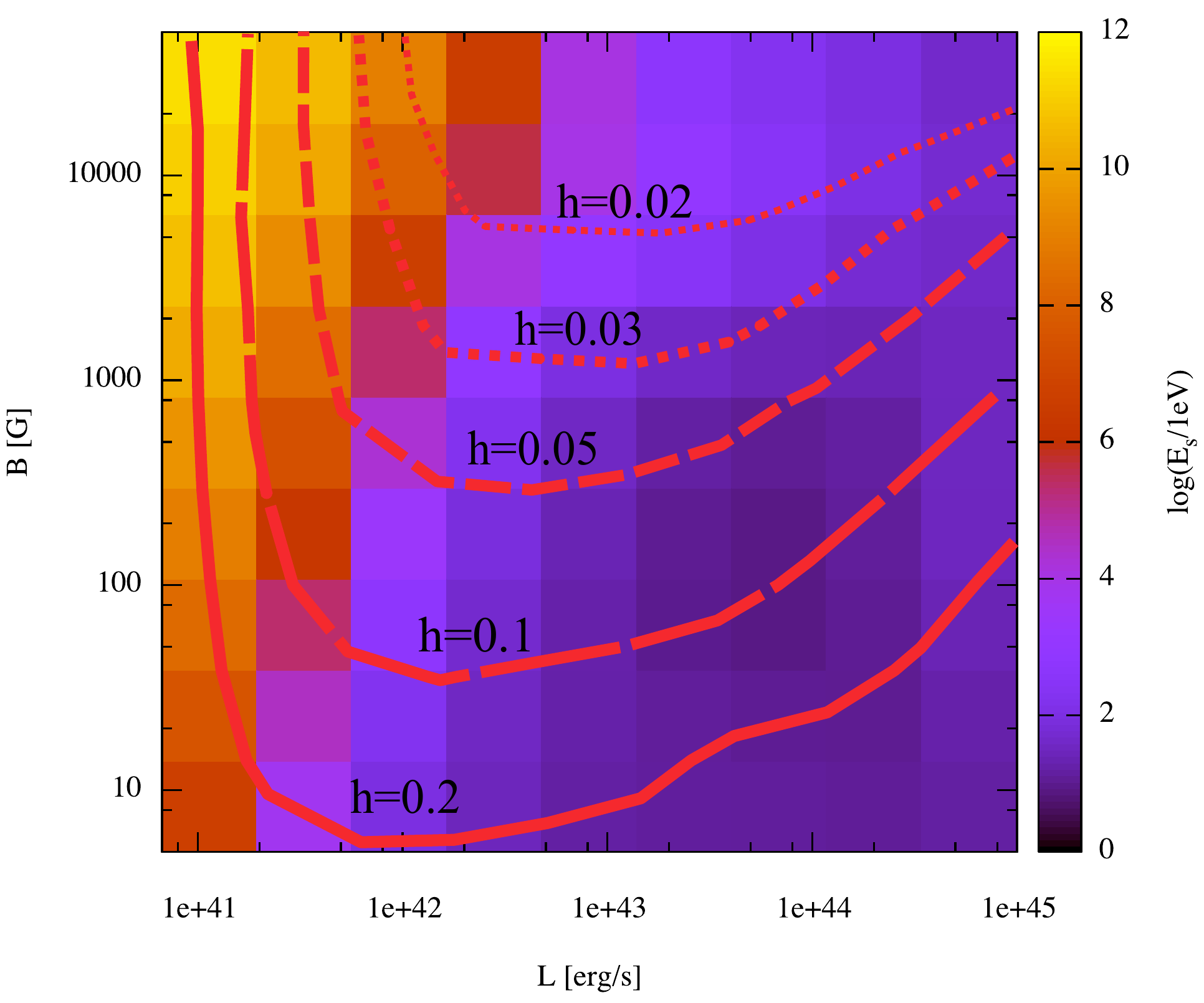}
 \caption{Maximal energies of synchrotron / curvature radiation photons produced by electrons in the gap, as a function of $L,B$. The black hole mass is $M=3\times 10^9M_\odot$. }
 \label{fig:synch_energy}
 \end{figure}

Higher energies of electrons lead to the higher energies of the synchrotron / curvature photons from these electrons. This is shown in Fig. \ref{fig:synch_energy}. One could notice that the energy of the synchrotron / curvature emission varies in a very broad range, from the ultraviolet / soft  X-ray up to the TeV \gr\ band. The highest energies of synchrotron / curvature photons violate the so-called "self-regulated cut-off" of the synchrotron spectrum, $E_s\lesssim m_e/\alpha$, where $\alpha$ is the fine structure constant \citep{aharonian_book}. This is because the acceleration proceeds in a large scale electric field aligned with the particle velocity. In such situation the acceleration rate is higher than $\sim eB_\bot$ assumed for the calculation of the "self-regulated" cut-off. 

\begin{figure}
\includegraphics[width=\linewidth]{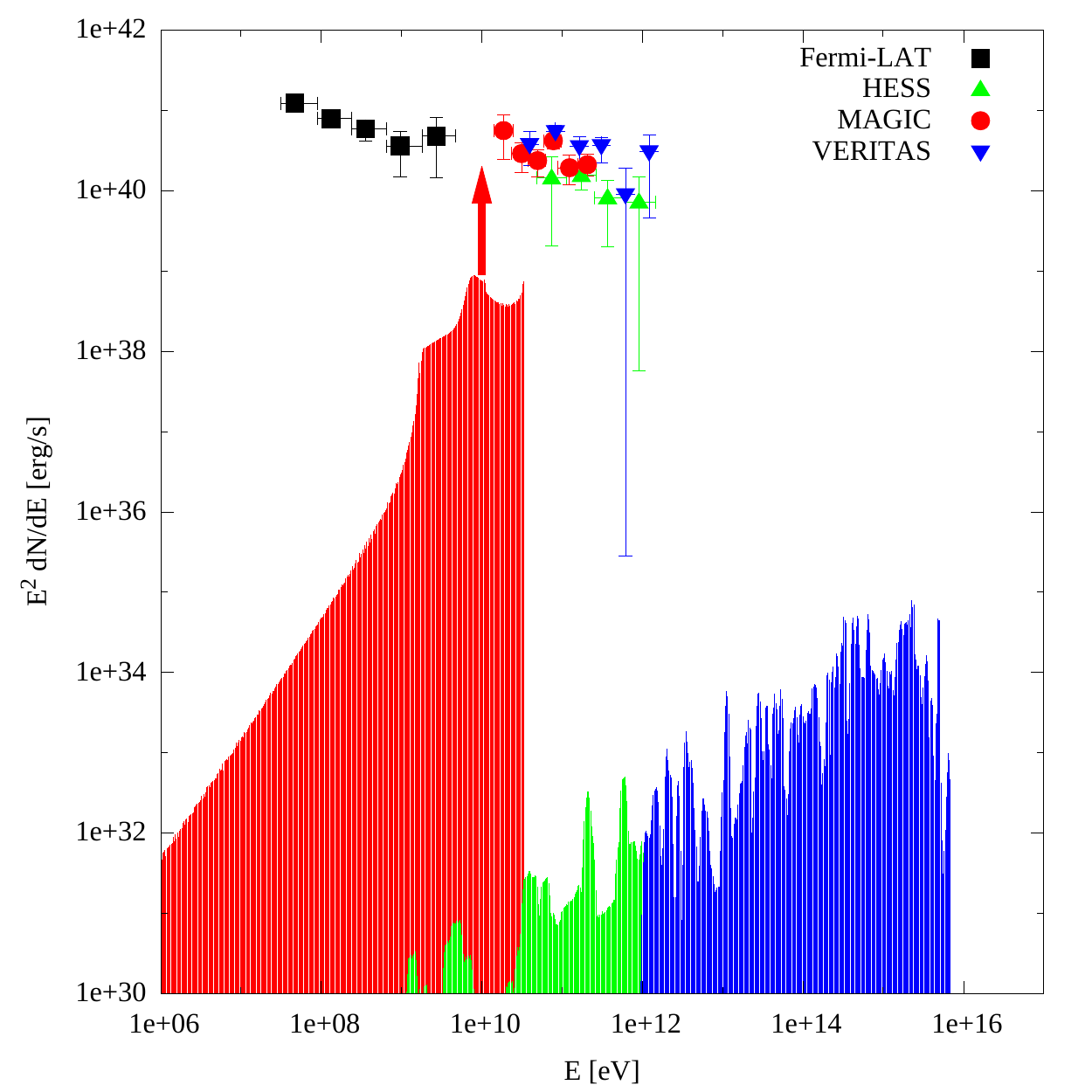}
\includegraphics[width=\linewidth]{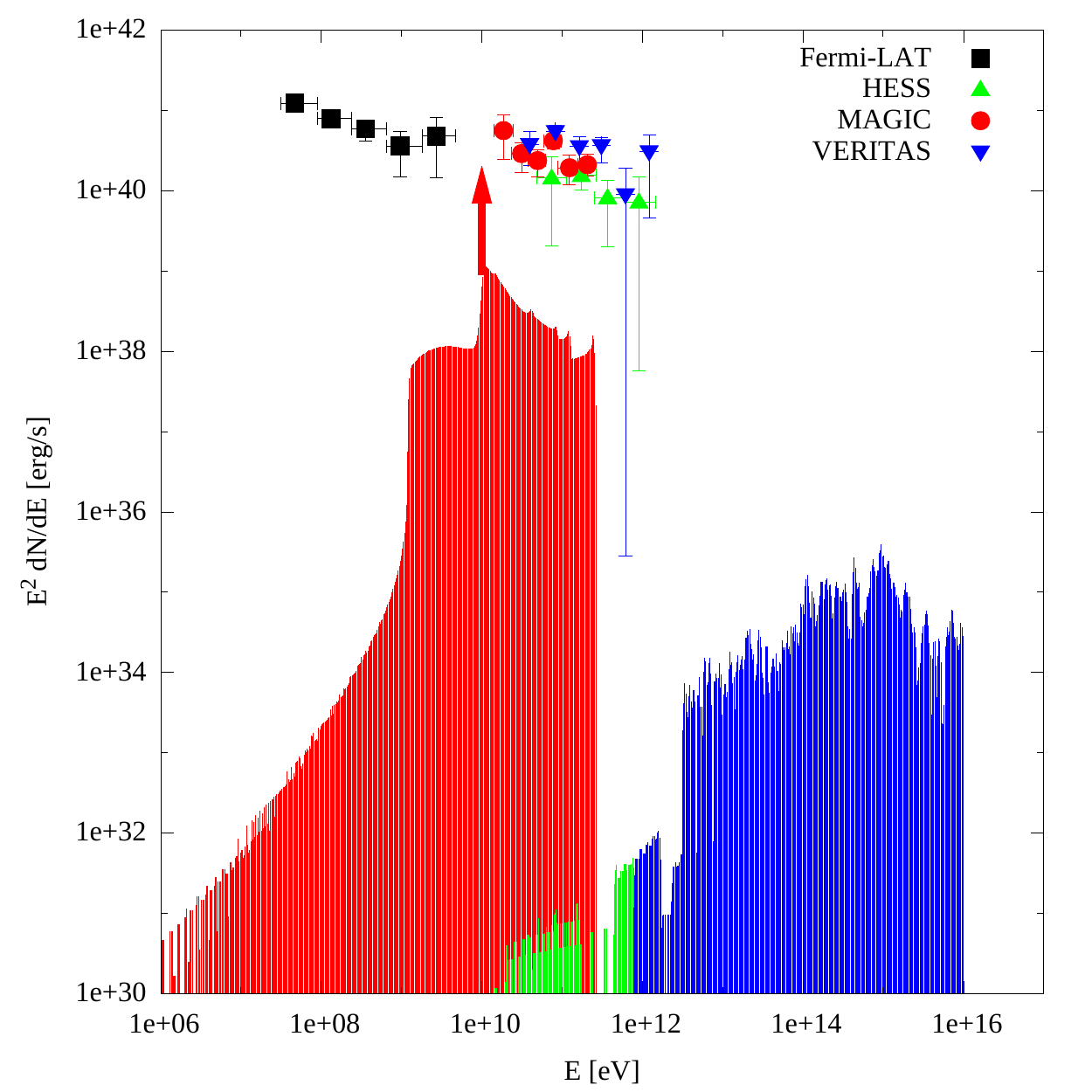}
\caption{Spectra of emission from electrons in the gap in the split monopole magnetosphere in a low-luminosity RIAF with $L=8\times 10^{40}$~erg/s around a black hole of the mass $M=3\times 10^9M_\odot$ and magnetic field $B=10^3$~G (top) and $B=10^4$~G (bottom). Red color shows synchrotron/curvature emission,  blue and green show the inverse compton emission. Green color corresponds to the photons which escape from the source, blue shows the photons absorbed via pair production. Red arrow shows the normalization uncertainty.}
\label{fig:spectrum_low}
\end{figure}

\begin{figure}
\includegraphics[width=\linewidth]{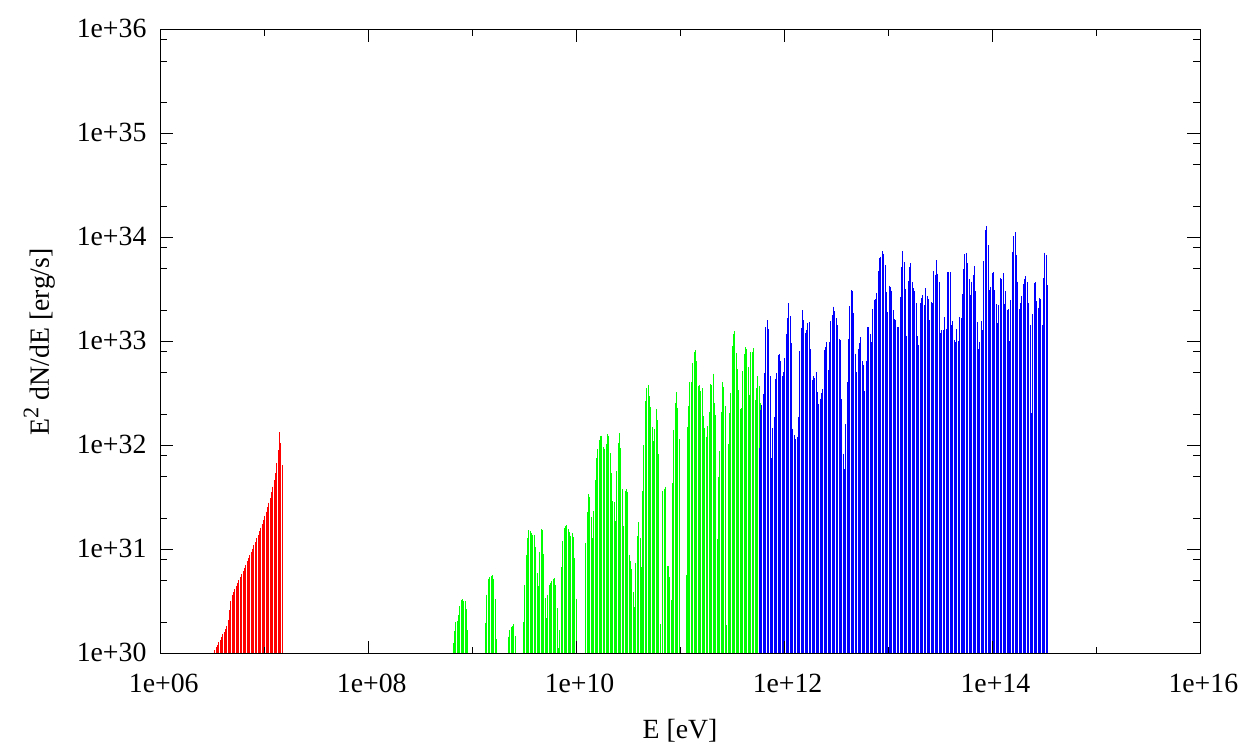}
\includegraphics[width=\linewidth]{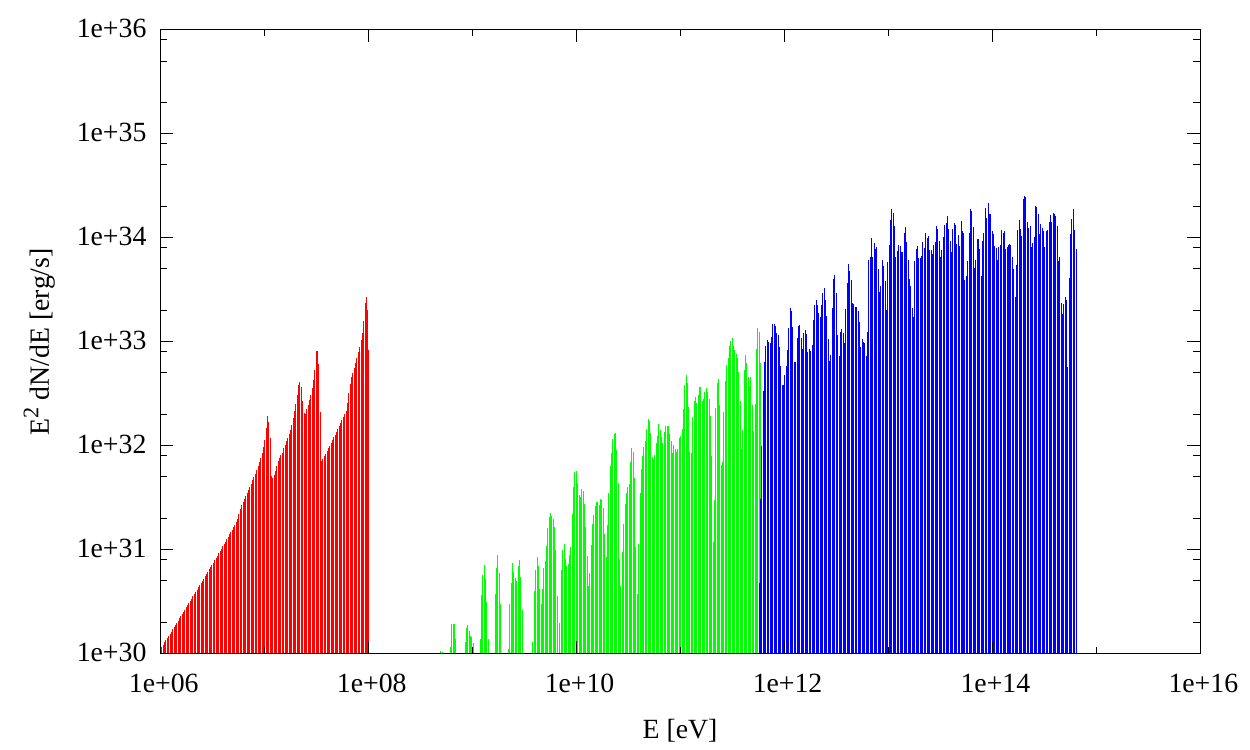}
\caption{Same as in Fig. \ref{fig:spectrum_low} but for the intermediate luminosity RIAF $L=5\times 10^{41}$~erg/s around $M=3\times 10^9M_\odot$ black hole in magnetic field $B=10^3$~G (top) and $B=10^4$~G (bottom). }
\label{fig:spectrum_mid}
\end{figure}

\begin{figure}
\includegraphics[width=\linewidth]{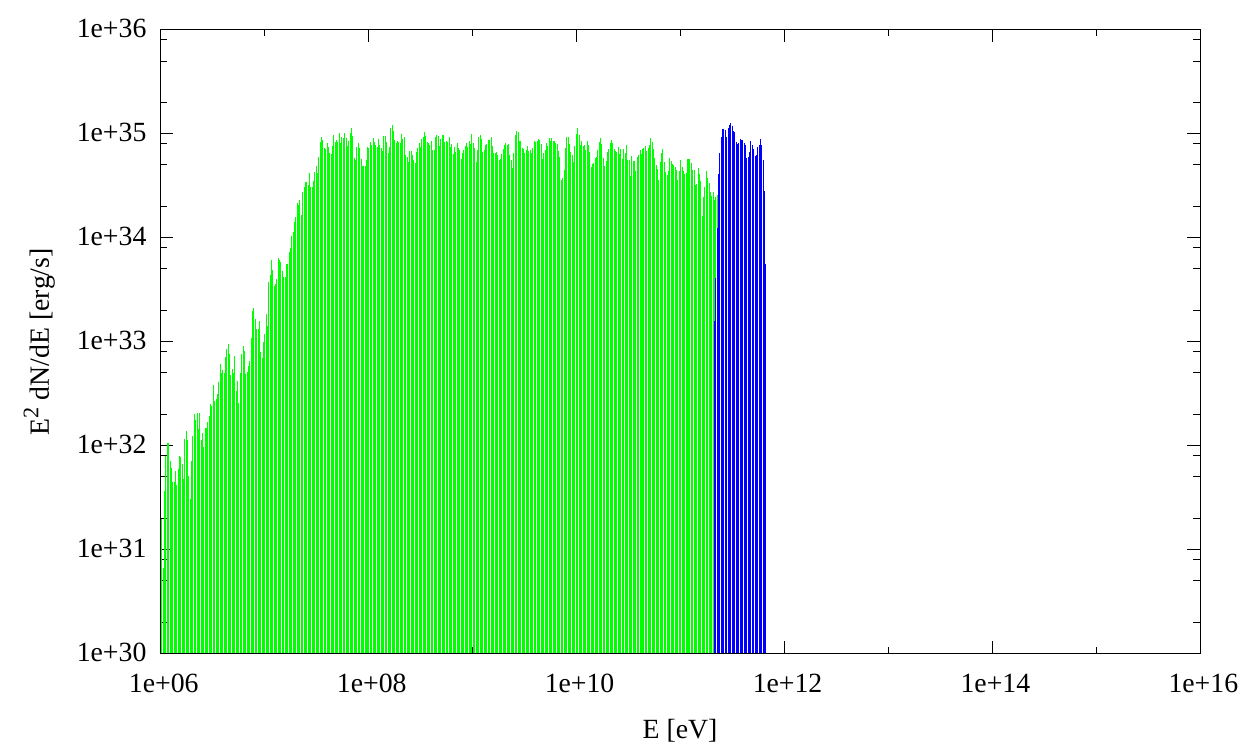}
\includegraphics[width=\linewidth]{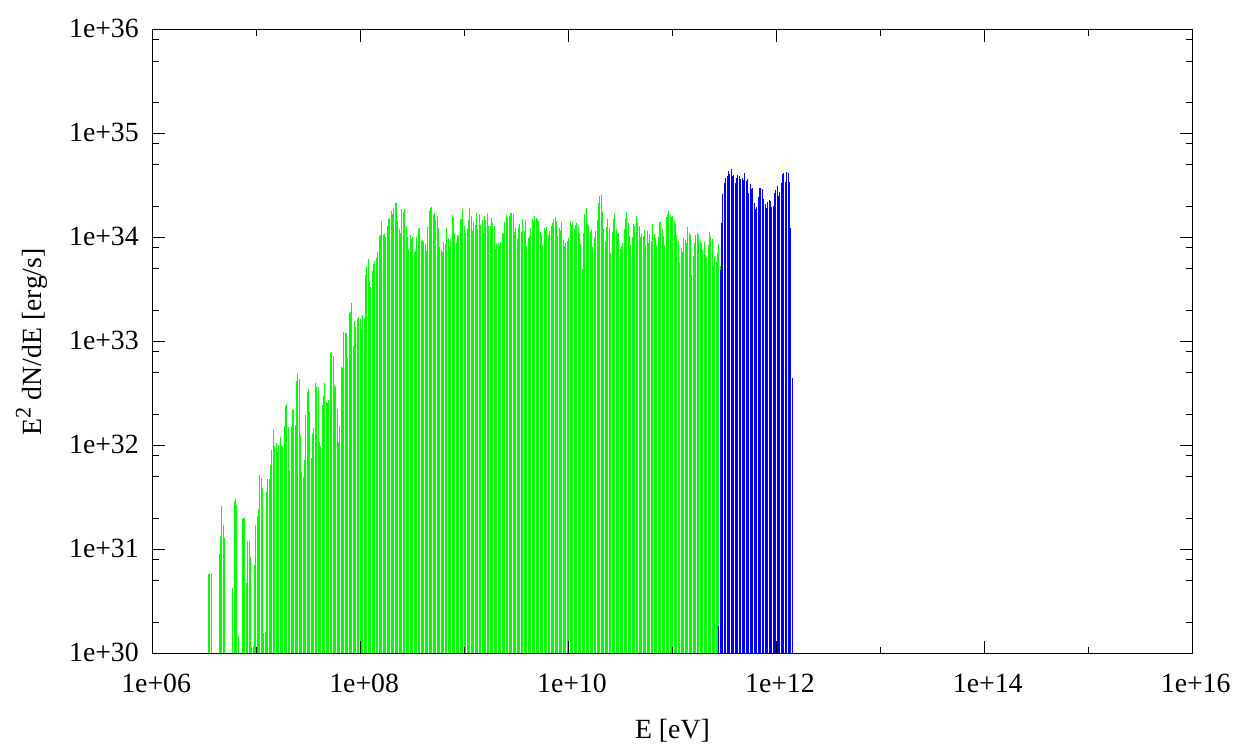}
\caption{Same as in Fig. \ref{fig:spectrum_low} but for the high-luminosity  RIAFs $L=10^{44}$~erg/s and magnetic field $B=100$~G (top) and $L=10^{43}$~erg/s and magnetic field $B=10^2$~G (bottom). The black hole mass is   $M=3\times 10^9M_\odot$ . }
\label{fig:spectrum_high}
\end{figure}

Pair production process limits the energy of inverse Compton photons to $E_{ic}\sim E_{\gamma,thr}$ independently of parameters of the system. This is clear from Figs. \ref{fig:spectrum_low}, \ref{fig:spectrum_mid}, \ref{fig:spectrum_high} showing the spectra of emission from electrons in the low, intermediate and high-luminosity systems. In all the spectra, the maximal energies of \gr s escaping to infinity are $E\sim 10^{12}$~eV, which corresponds to the redshifted energies of photons with energies comparable to the pair production threshold. 

The spectra of emission from the low, intermediate and high-luminosity RIAFs have qualitatively different appearance. 
The low-luminosity RIAF spectra shown in Fig. \ref{fig:spectrum_low}, the dominant flux contribution is from the synchrotron / curvature component which is sharply peaked in the 1-100~GeV range. The inverse Compton emission is largely subdominant and is perhaps barely detectable, provided that its flux is some 3 orders of magnitude below the GeV component flux. The GeV bump in the spectrum of the synchrotron / curvature emission is rather broad, spanning several decades in energy.
{The normalization of the gap spectra is defined by the maximum (Goldreich-Julian) density of the radiating electrons in the gap. Though the gap luminosity is uncertain and vary with the changes of parameters (\ref{fL}).} 

{One of the examples of the low-luminosity RIAF ($L\sim10^{40}-10^{41}\mbox{erg/s}$) is M87 (\cite{Neronov_and_Aharonian}, \cite{tchekhovskoy}).}
{The characteristic size of the infrared source could be estimated as $\sim 10R_H$. The direct mid-infrared observations of M87 \citet{Perlman_2007} \citet{Antonucci} resolves the central source within $\sim 10 \mathrm{pc} \sim 10^4R_{H}$. On the other side, millimeter wavelength observations \citet{Doeleman_2012} limits its size within horizon scale $R_H$. }  
{We compare the calculated spectra with M87 high energy observations (Fig. \ref{fig:spectrum_low}) (\cite{Fermi} \cite{HESS} \cite{VERITAS} \cite{MAGIC}). }  

In the higher luminosity RIAFs, the synchrotron / curvature component becomes sub-dominant. Its spectrum shifts to lower energies and becomes more peaked toward the higher energy. Overall, the synchrotron / curvature emission could appear as a narrow feature in the observed spectra of X-ray / soft \gr\ emission from intermediate RIAF luminosity sources, see Fig. \ref{fig:spectrum_mid}. To the contrary, the dominant very-high-energy inverse Compton emission in these sources has a very characteristic observational appearance: it is a relatively hard spectrum with the slope $dN/dE\propto E^{-\Gamma}$ with $\Gamma\sim 1.5$ and a abrupt ("super-exponential") cut-off at the energy of the threshold of the pair production. In a realistic situation, the cut-offf could be sharper or smoother depending on the details of the RIAF spectrum in the source, but the strong imprint of the pair production on the spectrum of emission from the vacuum gap is a generic feature of the VHE spectrum. 

 Spectra of the high-luminosity RIAF also have a strong imprint of the pair production, see Fig. \ref{fig:spectrum_high}. In these spectra, the synchrotron / curvature component might be not detectable at all, until the magnetic field is extremely high. Besides, strong radiation drag on the accelerated electrons produces a soft low-energy tail in the spectrum of the inverse Compton emission component.  
   
   

\begin{figure}
\includegraphics[width=\linewidth]{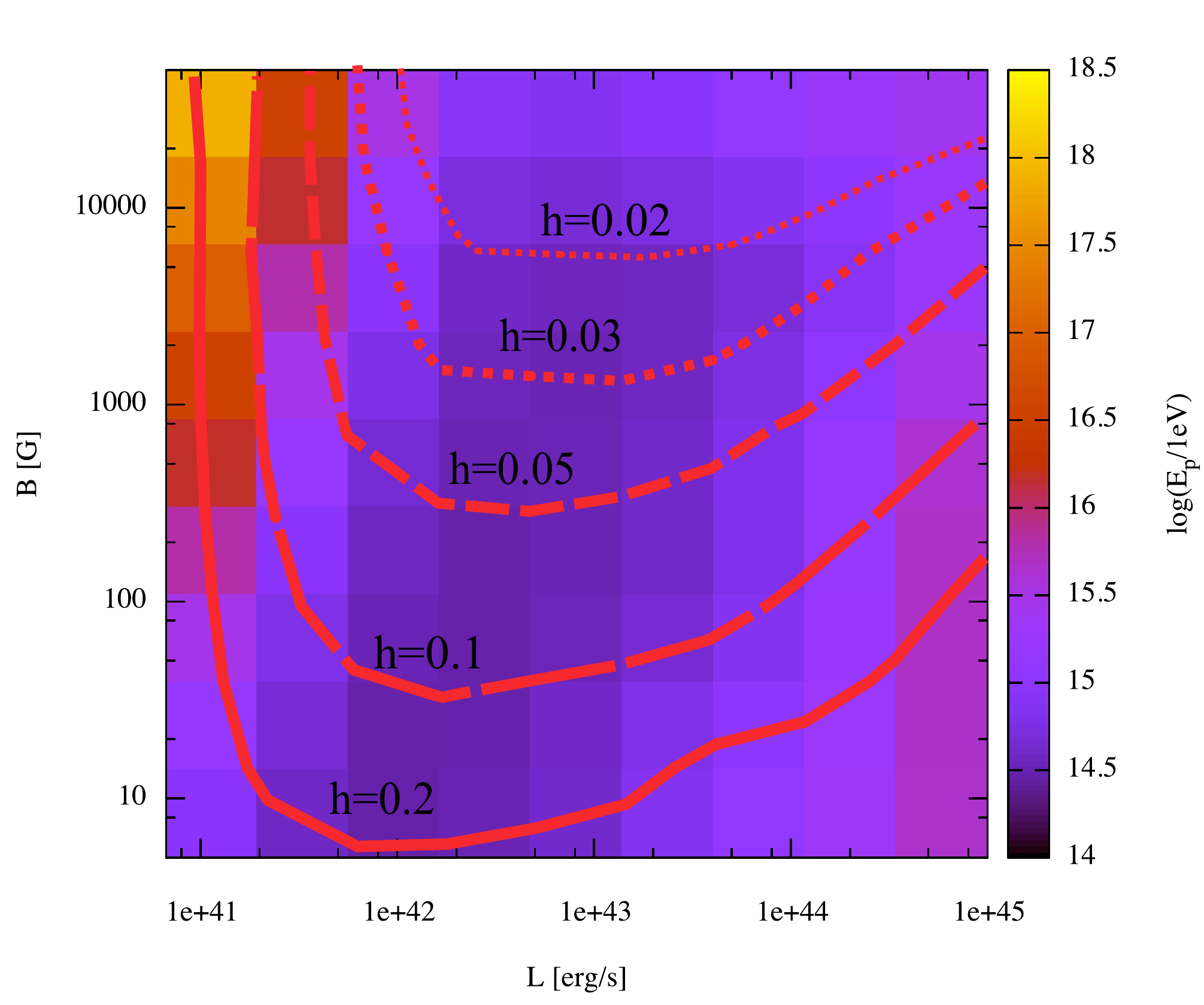}
\caption{Maximal energies of protons ejected from the gap. Black hole mass is $M=3\times 10^{9}M_\odot$. }
\label{fig:proton_energies}
\end{figure}

The maximum achievable energies protons, passing through the whole available potential difference in the gap do not depend strongly on the coordinate $\theta$. We can use the energy at the reference latitude $\theta=\pi/4$ as the characteristic values. The dependence of the proton maximal energies on $L$ and $B$ is shown in Fig. \ref{fig:proton_energies}.  In accordance with the analytical estimates, the maximum protons energies are not lower then ${\cal{E}}_p^{max}\geq10^{14}\mathrm{eV}$, where $10^{14}\mathrm{eV}$ is an energy, corresponding to the line $h_{*,acc}=\lambda_{\gamma\gamma}$. This minimal energy is reached exactly on this line in the $L,B$ diagram.  
    
In accordance with analytical estimates, in our numerical calculations we find that only energies of about ${\cal{E}}_p\sim 10^{17}-10^{18}$  are reachable in the gap in the magnetosphere of  black holes with masses in the range $M=10^5 - 10^9 M_\odot$  with magnetic field up to $B\sim 10^4$~G in low-luminosity RIAFs. This maximum energy is reached in the widest possible gaps and highest possible magnetic fields, in the low-luminosity RIAF regime. Our calculations do not extend to the situations in which UHECR production in black hole magnetospheres is possible. This would require still lower luminosity RIAF's where the gap height could become comparable to the size of the black hole (in other words, where the force-free magnetosphere does not form at all).

\section{Discussion and conclusions}    
\label{sec:discussion}  

In a realistic situation, the spectrum of electromagnetic emission from the vacuum gap is superimposed onto the spectrum of emission from the inner part of the FR I / BL Lac jet. The jet spectrum is typically represented by the broad synchrotron and inverse Compton components, which could be modelled as log-parabolas or cut-off powerlaws. Taking into account the similar appealrance of the gap and jet emission spectra, it might be challenging to prove / falsify the existence of the gap contribution to the overall spectra of the nuclear emission of the FR I galaxies and of the BL Lacs. 

In this respect, it is important to identify the distinguishing feature of the gap emission spectrum. This could be done for the low- , intermediate-  and high-luminosity RIAF cases based on Figs. \ref{fig:spectrum_low}, \ref{fig:spectrum_mid} and \ref{fig:spectrum_high}. 

The high- and intermediate luminosity RIAF cases are characterised by the strong inverse Compton component reaching the TeV band and having a sharp cut-off at the energy of the threshold of the pair production on the RIAF synchrotron photons. Presence of the shape cut-off is well understood qualitatively and appears to be the "distinguishing feature" of emission from the gap. The details of the spectral shape of the cut-off is determined by the details of the spectrum of the synchrotron component of RIAF. If this component is observed in the infrared band, the shape of the cut-off of the inverse Compton emission from the gap could be predicted. Observations of the FR I galaxies with identifiable synchrotron component of the RIAF and with the spectra extending into the very-high-energy (VHE)  range would provide a crucial test for the  gap emission model. The test will be the detection / non-detection of a VHE spectral component  component with the sharp cut-off with the shape determined by the shape of the RIAF spectral component. 

The sharply cut-off inverse Compton component in the interimdiate luminosity RIAF case also has a hard spectrum below the cut-off. The slope of the spectrum could be harder than the conventional slope $\Gamma\simeq 1.5$ expected from the $E^{-2}$ type distribution of electrons in the Thomson regime. Presence of the gap emission could then potentially provide an explanation to the hard spectra of VHE \gr\ emission from BL Lacs, such as the flaring spectrum of  Mrk 501 in 2009 \cite{neronov_hard}. Actually, the hard slope of the spectrum, combined with the sharp high-energy cut-off could make the gap emission appearing as a sharp, almost line-like (within the energy resolution of the \gr\ telescope) feature on top of a broad spectrum of inverse compton emission from the jet. 

The sharp VHE \gr\ band spectral feature is accompanied by a comparably sharp (or even sharper) feature at lower energies (soft \gr\ band) produced by the synchrotron / curvature emission from  electrons accelerated in the gap. This feature is also potentially detectable with the telescope sensitive in the MeV-GeV band, see Fig. \ref{fig:spectrum_mid}.

This sharp spectral feature in the VHE part of the spectrum is perhaps not detectable in the case of the emission from the gap embedded into low-luminosity RIAF. In this case only synchrotron / curvature emission is detectable. It appears as a hard spectrum / high-energy cut-off feature in the 1-100 GeV band and could also be readily distinguished from the broad inverse Compton emission from the jet, which usually dominates the spectrum of FR I galaxies and BL Lacs in this energy range, see Fig. \ref{fig:spectrum_low}. 

Overall, the observational appearance of the gaps in low / interimdiate / high luminosity RIAFs is always in the form of the sharp spectral features appearing on top of the broader / smoother  jet spectra. Improving quality of the \gr\ data expected with CTA should allow identification of these sharp features in the high-statistics spectral measurements of the spectra of BL Lacs and FR I galaxies. 
The data of Fermi/LAT telescope could be mined for the search of possibly transient narrow spectral components in the spectra of the LAT-detected sources from these source classes. 

\section*{Acknowledgements}
{We thank V.S. Beskin and D. Malyshev for the discussions. The work of K.P. was supported by the Russian Science Foundation grant 14-12-01340. K.P. acknowledges the Swiss Government Scholarship.
}


\end{document}